\documentclass[12pt]{article}
\usepackage{amssymb,amsmath}
\usepackage[dvips]{graphicx}
\usepackage{psfrag,epic,curves}

\title{Stroboscopic Variation Measurement}

\author{
  S.L.Danilishin\footnote{E-mail: stefan@hbar.phys.msu.ru},
  F.Ya.Khalili\footnote{E-mail: farid@mol.phys.msu.su}, \\
  \it Dept. of Physics, Moscow State University, Moscow 199899, Russia
}

\date{}

\begin{document}

\maketitle 

\begin{abstract}

A new procedure of the linear position measurement which allows to
obtain sensitivity better than the Standard Quantum Limit and
close to the Energetic Quantum Limit is proposed and analyzed in
detail. Proposed method is based on the principles of stroboscopic
quantum measurement and variation quantum measurement and allows
to avoid main disadvantages of both these procedures. This method
can be considered as a good candidate for use as a local position
meter in the ``intracavity'' topologies of the laser
gravitational-wave antennae.

\end{abstract}

\section{Introduction}

It has been known for many years that sensitivity of the weak force detectors
based on ``ordinary'' position meters is limited by the Standard Quantum
Limit \cite{67a1eBr} (see discussion on ``ordinary'' and other position
meters in \cite{00a1BrGoKhTh}). On the other hand, the scientific society
faced with the necessity to overcome this limit due to requirements of the
gravitational-wave experiments \cite{300years}. In order to detect
gravitational force producing small relative displacement of the antenna's
test masses, it is necessary to use methods that allow to overcome this
fundamental limit of sensitivity. There are two promising measurement
techniques that can be applied to solve this problem. These are stroboscopic
measurement and variation measurement methods.

Stroboscopic measurement \cite{78a1eBrKhVo} is the technique based on the
fact that harmonic oscillator momentum perturbation at $t=0$ does not
influence its coordinate at $t+n\pi/\omega_m$, where $\omega_m$ is the the
oscillator's eigenfrequency and $n$ is an integer. In other words, the back
action noise does not influence the measurement result if instantaneous
measurements are being made at regular time intervals of $\Delta t=\pi
n/\omega_m$. Using the stroboscopic measurement it is possible to detect
force

\begin{equation}\label{strob}
  F \simeq F_\mathrm{SQL}\sqrt{\omega_m\theta} \,,
\end{equation}
where $F_\mathrm{SQL}$ is the value of force corresponding to the
SQL and $\theta$ is the duration of each measurement.

Variation measurement technique can be divided into two main classes: the
time-domain variation measurement \cite{95a1VyZu, 98a1Vy} and spectral
variation measurement \cite{02a1KiLeMaThVy}. Time-domain variation
measurement allows to eliminate the back action noise from the meter's output
signal by introducing time-dependent cross-correlation between the meter's
measurement and back-action noise. In the particular case of interferometric
position meters being used in gravitational-wave detectors these noises
correspond to phase and power shot noises of the optical pumping
correspondingly, and this correlation can be introduced by modulation of the
local oscillator phase. In the spectral variation measurement
frequency-dependent cross-correlation is used.

Unfortunately, all these methods have their own disadvantages. In
order to achieve sensitivity substantially better than the SQL
using interferometric position meter in stroboscopic regime it is
necessary to use short pulses of optical pumping with high power
$W=\pi W_0/\omega_m\theta \gg W_0$, where $W_0$ is the mean power.
Moreover, as $\theta\ll\omega_m^{-1}$ the meter's output signal
contains additional noise due to brownian fluctuations of
high-frequency internal modes of the test body. Estimates show
that in the real gravitational-wave antennae this noise may be
comparable or even larger than the fundamental quantum noise.
Another disadvantage of the stroboscopic measurement is that it
can be applied only on meters with the harmonic oscillator for the
probe mass. At the same time only free masses are used for the
test object in the contemporary laser gravitational wave antennae.

On the other hand, in the case of the original form of time-domain variation
measurement it is necessary to know shape and arrival time of the signal to
be detected. As for the gravitational wave astronomy where the SQL overcoming
is crucial, this procedure can not be applied because arrival time as well as
exact shape of the signal are unknown. Spectral variation measurement is free
from this disadvantage, but in the case of interferometric position meters it
requires to use additional optical devices with bandwidth comparable with
bandwidth of the main interferometer. For example, in the case of LIGO
gravitational-wave observatory it is proposed to use additional Fabry-Perot
cavities with the length comparable with the length of the main cavities of
the antenna \cite{02a1KiLeMaThVy}.

In the articles \cite{98a1Vy, 00a1DaKhVy} the modified version of the
time-domain variation measurement which allows to circumvent these
disadvantages had been considered. This ``discrete sampling variation
measurement'' (DSVM) is based on approximation of the signal by series of
short rectangular ``slices'' with duration $\tau<\pi/\omega_{\rm max}$, where
$\omega_{\rm max}$ is the upper signal frequency, and periodical applying of the
variation measurement procedure fitted for such a rectangular pulse. In
accordance with sampling theorem signal shape can be completely restored
using signal values on consequent time intervals $t_n=n\tau$ as sampling
coefficients.

In this article we suggest new procedure which combines
time-dependent pumping of the stroboscopic measurement, the meter
noises time-dependent cross-correlation used in variation
measurement, and the discrete sampling method. We propose to call
this method ``stroboscopic-variation measurement'' (SVM). We will
show that such a combined procedure allows to obtain sensitivity
several times better than the pure DSVM procedure. This
sensitivity can be close to traditional forms of the variation
measurement and also close to the Energetic Quantum Limit
\cite{92BookBrKh,00p1BrGoKhTh}, which defines the ultimate
sensitivity that can be achieved at given meter's energy.

In the next section the idealized version of the SVM procedure
based on the the instant position measurements is considered. In
the section \ref{SVM} we show that rigorous optimization of the
variation measurements gives this procedure as a result. In the
section \ref{FinitePulse} and in the appendix \ref{app:A} we
consider more realistic version of the SVM procedure based on
position measurements with finite duration. In the appendix
\ref{App:B} the Energetic Quantum Limit for linear position meters
is considered in detail and it is also shown that sensitivity of
the spectral variation measurement is equal to this limit.
Appendix \ref{App:C} devoted to the explicit solution of the
optimization problem that arises in section \ref{SVM}.

\section{The idea of the SVM procedure}\label{idea}

Consider a slightly modified version of the triple measurement
procedure discussed in the article \cite{Likbez2001}. Let constant
force $F$ with duration $\tau$ to act on a harmonic test
oscillator with eigenfrequency $\omega_m$.  In order to detect
this force three instant measurements of the oscillator's position
are performed at the moments $t=0,\,t=\tau/2$, and $t=\tau$. The
results of these measurements can be presented as

\begin{equation}
  \tilde x_j = x_j + x_{j\,\rm fluct} \,,
\end{equation}
where $x_j\ (j=1,2,3)$ are the ``actual'' values of the test
object position at these moments of time,

\begin{subequations}
  \begin{align}
    x_2 &= x_1\cos\frac{\omega_m\tau}{2}
      + \frac{p_1+p_{1\,\rm fluct}}{m\omega_m}\,\sin\frac{\omega_m\tau}{2}
      + \frac{F}{m\omega_m^2}\,\left(1-\cos\frac{\omega_m\tau}{2}\right)\,,\\
    x_3 &= x_1\cos\omega_m\tau
      + \frac{p_1+p_{1\,\rm fluct}}{m\omega_m}\,\sin\omega_m\tau
      + \frac{p_{2\,\rm fluct}}{m\omega_m}\,\sin\frac{\omega_m\tau}{2}
      + \frac{F}{m\omega_m^2}\,(1-\cos\omega_m\tau) \,,
  \end{align}
\end{subequations}
where $p_1$ is the initial momentum of the test object and
$p_{1,2\,\rm fluct}$ are the test object momentum perturbations
during the first and the second measurement correspondingly.

Estimate of the force $F$ which does not depend on the initial state of the
test object can be obtained using the formula

\begin{equation}\label{tilde_F}
  \tilde F
  = \frac{m\omega_m^2}{2\left(1-\cos\frac{\omega_m\tau}{2}\right)}\,
    \left(
      \tilde x_1 - 2\tilde x_2\cos\frac{\omega_m\tau}{2} + \tilde x_3
  \right) \,.
\end{equation}
Uncertainty of this value $\Delta F$ depends on the  mean square
errors for each of the measurement $\Delta_{x\,j}$, corresponding
mean square perturbations of the test object momentum
$\Delta_{p\,j}$, and the cross-correlations between the
measurement errors and perturbations $\Delta_{xp\,j}$. They must
satisfy the Heisenberg relation

\begin{equation}\label{Heizenberg}
  \Delta_{x\,j}^2\Delta_{p\,j}^2 - \Delta_{xp\,j}^2 = \frac{\hbar^2}{4} \,.
\end{equation}
The experimentalist should optimize measurement by minimizing value of
$\Delta F$. Taking into account formula (\ref{Heizenberg}) it is easy to show
that optimal values of cross-correlation are equal to

\begin{align}\label{Delta_opt}
  \Delta_{xp\,1} &= \Delta_{xp\,3} = 0 \,, &
  \Delta_{xp\,2} &= \frac{\Delta_{p\,2}^2}{2m\omega_m}\,
    \tan\frac{\omega_m\tau}{2} \,.
\end{align}
and in this case

\begin{equation}\label{Delta_F_opt}
  (\Delta F)^2
  = \frac{m^2\omega_m^4}{4\left(1-\cos\frac{\omega_m\tau}{2}\right)^2}\,
    \left(
      \frac{\hbar^2}{4\Delta_{p\,1}^2}
      + \frac{\hbar^2}{\Delta_{p\,2}^2}\,\cos^2\frac{\omega_m\tau}{2}
      + \frac{\hbar^2}{4\Delta_{p\,3}^2}
    \right) \,.
\end{equation}

Described procedure is very close to the variation measurement.
Really, we eliminated the back-action term in the expression
(\ref{Delta_F_opt}) using cross-correlation between $x_{2\,\rm
fluct}$ and $p_{2\,\rm fluct}$ (see formulae (\ref{Delta_opt})).
In such a procedure it is possible to obtain, in principle,
arbitrary high precision by increasing the values of
$\Delta_{p\,j}$.

On the other hand, if $\omega_m\tau=\pi n$ then the term corresponding to the
second measurement in the formulae (\ref{tilde_F},\ref{Delta_F_opt})
vanishes. It means that only two measurements are necessary in this case, the
first and the third ones, which correspond to the stroboscopic procedure.
Therefore stroboscopic measurement which uses  time-dependent pumping and
variation measurement which uses time-dependent cross-correlation of noises
should be considered as particular cases of ``stroboscopic-variation''
procedure described here.

In the real experiment the values of $\Delta_{p\,j}$ can not be
arbitrary high due to energy limitations in the meter. Therefore
value (\ref{Delta_F_opt}) can be further optimized taking into
account additional condition that

\begin{equation}\label{sumDeltaP}
  \Delta_{p\,1}^2 + \Delta_{p\,2}^2 + \Delta_{p\,3}^2
  = \overline{S}_F\tau \,,
\end{equation}
where $\overline{S}_F$ is some given value (sense of this peculiar
notation will be explained below too). In the case of the
interferometric position meter it is proportional to the optical
energy stored in the interferometer, averaged over the time
$\tau$.

It is easy to show that in the optimal case we will obtain the following
expression:

\begin{equation}\label{Delta_opt_opt}
  \Delta_{p\,1}^2 = \Delta_{p\,3}^2
   = \frac{1}{2}\,\left|\cos\frac{\omega_m\tau}{2}\right|\,\Delta_{p\,2}^2 \,.
\end{equation}

It is convenient to describe sensitivity by the meter effective
noise spectral density

\begin{equation}\label{DiscreteSpDensity}
  S_{\rm meter} = (\Delta F)^2\tau \,,
\end{equation}
as it was done in the article \cite{00a1DaKhVy}. If the conditions
(\ref{Delta_opt_opt}) are valid, then the sensitivity is defined by the
following formulae:

\begin{equation}\label{Delta_F_opt_opt}
  \frac{S_{\rm meter}}{S_\mathrm{EQL}} = \begin{cases}
    \left(\dfrac{\omega_m\tau}{\pi}\right)^4\cot^4\dfrac{\omega_m\tau}{4}\,,
      & \text{if\ } 0 \le \omega_m\tau \le \pi \,, \medskip \\
    \left(\dfrac{\omega_m\tau}{\pi}\right)^4
      & \text{if\ } \pi \le \omega_m\tau \le 2\pi \,,
  \end{cases}
\end{equation}
for the harmonic oscillator, and

\begin{equation}\label{Delta_F_opt_opt_fm}
  \frac{S_{\rm meter}}{S_\mathrm{EQL}} = \left(\frac{4}{\pi}\right)^4
\end{equation}
for the free mass ($\omega_m\to 0$),

Here

\begin{equation}\label{Delta_F_EQL}
  S_\mathrm{EQL} = \frac{\pi^4\hbar^2 m^2}{4\overline{S}_F\tau^4}
\end{equation}
is a convenient scale factor for the meter's noise equal to the Energetic
Quantum Limit in the free test mass case taken at frequency (see Appendix
\ref{App:B}).

It should be noted that $S_{\rm meter}=S_\mathrm{EQL}$ only in the
case of the pure stroboscopic measurement ($\omega_m\tau=\pi$),
and $S_{\rm meter}>S_\mathrm{EQL}$ for all other values of
$\omega_m\tau$ (see the lowest curve in the Fig.\,\ref{fig:4}).

\section{Linear position meter with time-dependent parameters}\label{SVM}

Consider now more realistic situation when the meter which can
continuously monitor position $x(t)$ of the test object (for
example, interferometric position meter) is used to detect signal
force $F_{\rm signal}(t)$. When the meter is turned on, its output
signal can be represented as

\begin{equation}\label{xsign}
  \tilde x(t) = \hat x_{init}(t) + \hat x_{\rm fluct}(t)
    + {\bf D}^{-1}[F_{\rm signal}(t)+\hat F_{\rm fluct}(t)] \,,
\end{equation}
where ${\bf D}$ is the linear differential operator describing the
dynamics of the test object\footnote{In the case of the harmonic
oscillator operator ${\bf D}$ can be written as$${\bf
D}=\frac{d^2}{dt^2}+\omega_0^2\,.$$}, $\hat x_{\rm fluct}(t)$ is
the noise added by the meter, $\hat F_{\rm fluct}(t)$ is the
back-action force, and $\hat x_{init}(t)$ is the ``initial''
position of the test object which describes its motion when signal
force and back action force of the meter are absent. It should be
noted that, whereas $\hat x_{\rm fluct}(t), \hat F_{\rm fluct}(t)$
and $\hat x_\mathrm{init}(t)$ are quantum-mechanical operators,
$\tilde x(t)$ is a classical observable (see discussion on
de-quantization in linear position meters in the article
\cite{Likbez2001}).

Spectral densities of these noises $S_x$ and $S_F$, and their
cross spectral density $S_{xF}$ satisfy the uncertainty relation

\begin{equation}\label{noises}
  S_x S_F - S_{xF}^2 = \frac{\hbar^2}{4} \,.
\end{equation}
Suppose that the experimentalist can change values of these
spectral densities within the boundaries defined by the condition
(\ref{noises}). If the interferometric position meter is used, the
experimentalist can change the ratio $S_F/S_x$ changing the
pumping power (as in the case of the stroboscopic measurement) and
also he can change the value of $S_{xF}$ by changing the phase of
the local oscillator (as in the case of the variational
measurement).

We suppose in this article for simplicity that noises $\hat F_{\rm
fluct}(t)$ and $\hat x_\mathrm{init}(t)$ can be regarded as
``white'' ones, and that the values of the spectral densities can
be changed instantly (without any inertia). It means that the
experimentalist have to be able to change the pumping energy in
the interferometric position meter very quickly compared to the
signal frequency. In the real experiments these approximations are
valid only if relaxation time of the interferometric meter
cavities is small compared to the signal characteristic time-scale
(see brief discussion on this topic in the conclusion).

On the first stage of signal processing information about the test
object initial conditions (term $x_{\rm init}(t)$) is being
excluded from the output signal $\tilde x(t)$ by applying operator
$\bf D$ to (\ref{xsign}). As a result one will obtain the
following expression for the estimation of the force $F(t)$:

\begin{equation}\label{outsign}
  \tilde F(t) = {\bf D}\tilde x(t) = F_{\rm signal}(t)
    + \hat F_{\rm fluct}(t) + {\bf D}\hat x_{\rm fluct}(t) \,.
\end{equation}
The next stage is the optimal estimation of the force $F(t)$
averaged over short time interval, as it has been suggested in the
DSVM method \cite{98a1Vy,00a1DaKhVy}. Mean-square error of
this estimation is equal to

\begin{equation}\label{measerr1}
  (\Delta F)^2 =\frac{\hbar^2}{4}\int\limits_{-\tau/2}^{\tau/2}
    \frac{({\bf D}v(t))^2}{S_F(t)}\,dt
  + \int\limits_{-\tau/2}^{\tau/2}S_F(t)[a(t){\bf D}v(t)+v(t)]^2\,dt,
\end{equation}
where $a = S_{xF}/S_F$ is the variational factor and $v(t)$ is the filter
function which must satisfy conditions

\begin{align}\label{bound_norm}
  v(t)\Bigr|_{t=\pm\tau/2} &= 0 \,, &
  \frac{dv(t)}{dt}\Bigr|_{t=\pm\tau/2} &= 0 \,, &
  \int\limits_{-\tau/2}^{\tau/2}v(t)\,dt &= 1 \,.
\end{align}
We also suppose that function $S_F(t)$ is limited by the condition

\begin{equation}\label{normS}
  \int\limits_{-\tau/2}^{\tau/2}S_F(t)\,dt =\overline S_F\tau\,.
\end{equation}
(compare with condition (\ref{sumDeltaP})).

One can readily see that functional (\ref{measerr1}) is a sum of
two non-negative items, and variational coefficient $a(t)$ appears
only in the second one which originates in the back-action of the
meter. It is evident that if variational factor $a(t)$ satisfies
the equation:

\begin{equation}\label{aeqn}
  a(t){\bf D}v(t)+v(t)=0\,
\end{equation}
then the second term vanishes and expression (\ref{measerr1})
takes form

\begin{equation}\label{measerr2}
  (\Delta F)^2 =\frac{\hbar^2}{4}\int\limits_{-\tau/2}^{\tau/2}\frac{({\bf
 D}v(t))^2}{S_F(t)}\,dt.
\end{equation}
Expressions (\ref{measerr2}), (\ref{bound_norm}), and (\ref{normS}) form the
optimization problem to be solved in order to find characteristics for
optimal signal processing.

It should be noted that the standard Lagrange optimization
procedure can not be used here, as function under integral in
(\ref{measerr2}) is not a continuously differentiable one, so the
Lagrange equations could not be solved. This problem can be solved
using the Liapunov's problems optimization theory (See
\cite{Galeev} for details). The explicit solution is presented in
Appendix \ref{App:C}, and here we will provide only the final
expression.

In the particular case of arbitrarily short pumping pulses the
optimal function $S_F(t)$ is equal to
\begin{equation}\label{OscSFOptimum}
  S_F(t)
  =
 \begin{cases}
  \dfrac{\overline{S}_F\tau}{
      2\left(1+\cos\frac{\omega_m\tau}{2}\right)}
    \Bigl[
      \delta(t-\tau/2) + 2\cos\frac{\omega_m\tau}{2}\,\delta(t)
      + \delta(t+\tau/2)
    \Bigr]\,, & \mbox{if }\omega_m\tau<\pi\,,\\
    \dfrac{\overline{S}_F\tau}{2}\left[\delta(t-\frac{\pi}{2\omega_m})+\delta(t+\frac{\pi}{2\omega_m})\right]\,,
    & \mbox{if }\omega_m\tau>\pi\,,
  \end{cases}
\end{equation}
The measurement error in this case is described by the formulae
(\ref{Delta_F_opt_opt},\ref{Delta_F_opt_opt_fm}).

So we have shown that the sequence of the three instant
measurements described in section \ref{idea} represents an optimal
procedure when the experimentalist is able to change freely the
values of $S_F,S_x$ and $S_{xF}$.

\section{Pumping pulses with finite duration}\label{FinitePulse}

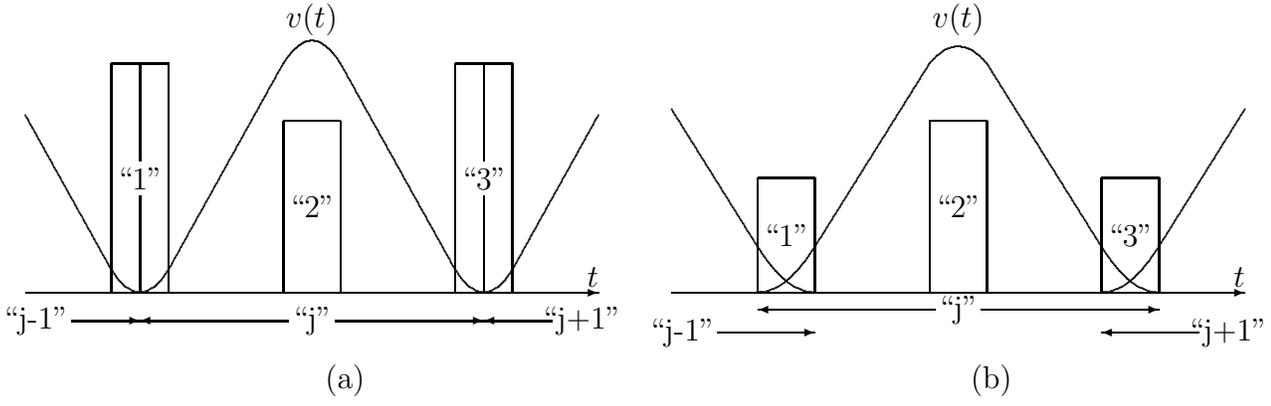
\begin{figure}

\begin{minipage}[c]{0.5\textwidth}

\begin{picture}(100,50)

\put(0,10){\vector(1,0){100}}\put(100,11){\makebox(0,0)[br]{$t$}}

\drawline(15,10)(15,50)(25,50)(25,10)
\drawline(20,10)(20,27)\drawline(20,33)(20,50)
\put(20,30){\makebox(0,0)[cc]{``1''}}

\drawline(45,10)(45,40)(55,40)(55,10)\put(50,25){\makebox(0,0)[cc]{``2''}}

\drawline(75,10)(75,50)(85,50)(85,10)
\drawline(80,10)(80,27)\drawline(80,33)(80,50)
\put(80,30){\makebox(0,0)[cc]{``3''}}

\curve(20,10,22.5,11,25,14)\curve(25,14,45,50)\curve(45,50,50,54,55,50)
\curve(55,50,75,14)\curve(75,14,77.5,11,80,10)
\put(50,55){\makebox(0,0)[cb]{$v(t)$}}

\curve(0,41,15,14)\curve(15,14,17.5,11,20,10)

\curve(80,10,82.5,11,85,14)\curve(85,14,100,41)

\put(8,5){\vector(1,0){12}}\put(-4,5){\makebox(0,0)[lc]{``j-1''}}

\put(46,5){\vector(-1,0){26}}\put(54,5){\vector(1,0){26}}
\put(50,5){\makebox(0,0)[cc]{``j''}}

\put(92,5){\vector(-1,0){12}}\put(104,5){\makebox(0,0)[rc]{``j+1''}}

\end{picture}
\centerline{(a)}
\end{minipage}
\begin{minipage}[c]{0.5\textwidth}

\begin{picture}(100,50)

\put(0,10){\vector(1,0){100}}\put(100,11){\makebox(0,0)[br]{$t$}}

\drawline(15,10)(15,30)(25,30)(25,10)\put(20,20){\makebox(0,0)[cc]{``1''}}

\drawline(45,10)(45,40)(55,40)(55,10)\put(50,25){\makebox(0,0)[cc]{``2''}}

\drawline(75,10)(75,30)(85,30)(85,10)\put(80,20){\makebox(0,0)[cc]{``3''}}

\curve(15,10,20,12,25,18)\curve(25,18,45,50)\curve(45,50,50,53,55,50)
\curve(55,50,75,18)\curve(75,18,80,12,85,10)
\put(50,55){\makebox(0,0)[cb]{$v(t)$}}

\curve(0,42,15,18)\curve(15,18,20,12,25,10)

\curve(75,10,80,12,85,18)\curve(85,18,100,42)

\put(8,3){\vector(1,0){17}}\put(-4,3){\makebox(0,0)[lc]{``j-1''}}

\put(46,7){\vector(-1,0){31}}\put(54,7){\vector(1,0){31}}
\put(50,7){\makebox(0,0)[cc]{``j''}}

\put(92,3){\vector(-1,0){17}}\put(104,3){\makebox(0,0)[rc]{``j+1''}}

\end{picture}
\centerline{(b)}
\end{minipage}
\caption{Different layouts of the pumping pulses}\label{fig:pulses}

\end{figure}

It is evident, of course, that $\delta$-like pumping pulses could
not be obtained in practice. Therefore we consider here pumping
pulses with finite duration.

In order to reconstruct the signal force shape, the sequence of
triple measurements described above have to be used. The
experimentalist has a choice here, whether to use completely
independent triads with each of them containing all three pulses
(see Fig.\,\ref{fig:pulses}(a)) or use overlapping triads with
common first and third pulses (see Fig.\,\ref{fig:pulses}(b)). The
second variant is more ``energy-saving'', but it does not permit
to use variational technique for the first and the third pulses.
It is possible to show that due to this reason its sensitivity is
limited by the same condition (\ref{strob}) as sensitivity of the
traditional stroboscopic procedure. Here we will consider the
first variant only.


In this case sensitivity is described by the formula

\begin{equation}\label{GenErr}
  S_{\rm meter}
  = \frac{\hbar^2 m^2\omega_m^4}{4\overline{S}_F(1 - a_1^2/a_2)}\,
  = \frac{(\omega_m\tau/\pi)^4}{1 - a_1^2/a_2}\,{S_\mathrm{EQL}} \,,
\end{equation}
where

\begin{equation}\label{GenCoeff}
  a_j = \frac1{\overline{S}_F\tau}
    \int_{-\tau/2}^{\tau/2} S_F(t)\cos^j\omega_m t\,dt\,,\mbox{ where
    }j=1,2\,.
\end{equation}
It depends on the shape of the pumping pulse, but it is evident
that in any case the larger is $S_F$ (\textit{i.e.} the pumping
power), the smaller is noise, and there are no absolute
limitations similar to the SQL or the formula (\ref{strob}) here.

Consider now quasi-optimal case when pumping pulses has small but
finite duration $\theta\ll\tau$. Suppose that $\theta$ is the
shortest pumping pulse duration that can be used in real
experiment. In any case duration $\theta$ should be large compared
to the interferometer's cavities relaxation time. Let also the
first and the third pulses to be divided between consequent
measurement cycles, as it is shown in Fig. \ref{fig:pulses} (a).
In this case function $S_F(t)$ can be presented as

\begin{equation}\label{QuasiOpt1}
\begin{array}{cc}
  S_F(t) = \overline{S}_F\tau
    \bigl[k\Delta(t+\tau/2)+(1-k)\Delta(t)+k\Delta(t-\tau/2)\bigr]
    \,,\\
    \\
    \mbox{ where }t\in\left[\tau\left(j-\frac12\right),\,\tau\left(j+\frac12\right)\right]\,.
\end{array}
\end{equation}
Here factor $k$ describes the relative energy of the pumping
pulses, $0\le k\le 1$, and $\Delta(t)$ is the narrow
$\delta$--like function that describes pumping pulse shape, and
let

\begin{align}
  \int\limits_{-\tau/2}^{\tau/2}\Delta(t)\,dt &= 1 \,, &
  \int\limits_{-\tau/2}^{\tau/2}|t|\Delta(t)\,dt &= \alpha_1\theta \,, &
  \int\limits_{-\tau/2}^{\tau/2}t^2\Delta(t)\,dt
    &= (\alpha_2+\alpha_1^2)\theta^2 \,,
\end{align}
where $\alpha_1,\alpha_2$ are some numeric factors which depend on
the shape of the pulses.

Substituting these expression into the formula (\ref{GenErr}), neglecting the
terms of order of magnitude $\theta^3$ or higher and optimizing the result
with respect to $k$, we will obtain that

\begin{equation}\label{GenErrFinal}
  \frac{S_{\rm meter}}{S_\mathrm{EQL}} \approx
  \begin{cases}
    \left(\dfrac{\omega_m\tau}{\pi}\right)^4\cot^4\dfrac{\omega_m T}{4}\left[
      1 + 2\dfrac{
        \alpha_1^2\cos\frac{\omega_m T}{2}
        - 2\alpha_2(1+\cos\frac{\omega_m T}{2})
      }{\sin^2\frac{\omega_m T}{2}}\,
      \omega_m^2\theta^2\right]\, &
      \text{if\ } \omega_mT < f(\theta) \,, \\
    \left(\dfrac{\omega_m\tau}{\pi}\right)^4\, &
      \text{if\ } \omega_m T > f(\theta) \,,
  \end{cases}
\end{equation}
where $T=\tau-2\alpha_1\theta$ and $f(\theta) \approx
\pi+(\alpha_1^2-\alpha_2)\omega_m^2\theta^2$.

These expressions correspond to the following optimal values of $k$:

\begin{equation}\label{Genk}
  k \approx \begin{cases}
    \dfrac{1}{2\cos^2\frac{\omega_mT}{4}}\left[
      1 + \left(
        \dfrac{\alpha_2}{\sin^2\frac{\omega_mT}{4}}
        - \dfrac{\alpha_1^2\cos\frac{\omega_mT}{2}}
            {4\cos^4\frac{\omega_mT}{4}}\,
      \right)\omega_m^2\theta^2
    \right]\, & \text{if\ } \omega_m T < f(\theta) \,, \smallskip \\
      1\,, & \mbox{if}\quad f(\theta)<\omega_mT<\pi\,,  \smallskip  \\
    \dfrac{1}{2\sin^2\frac{\omega_mT}{4}}\left[
      1 + \dfrac{\alpha_1^2\cos\frac{\omega_mT}{2}}
            {4\sin^2\frac{\omega_mT}{4}}\,\omega_m^2\theta^2
    \right]\, & \text{if\ } \omega_m T > \pi \,.
  \end{cases}
\end{equation}

In the appendix \ref{app:A}  rectangular pumping pulses with
arbitrary duration $\theta$ are considered and explicit formulae
for the measurement error are provided for this case. We suppose
rectangular pulses to be a good model for real pumping pulses as
the main disadvantage of square shaped function, that is infinite
derivative on edges, does not influence the final expression for
the measurement error, because due to (\ref{GenCoeff}) function
$S_F(t)$ that consists of rectangular pulses, is included to
$\Delta F$ as integrand only and the final expression does not
depend on its derivatives.

\section{Conclusion}

Stroboscopic variation measurement presented in this article allows to obtain
sensitivity better than the Standard Quantum Limit and close to the Energetic
Quantum Limit using time-dependent values of meter noises spectral densities.
In particular, if interferometric position meter is used then this procedure
can be implemented by using both pumping power and phase of the local
oscillator modulation. At the same time, it does not require pumping power to
be in a non-classical state.

It is important that the stroboscopic variation measurement does
not require information about the signal shape and arrival time
(as the usual variation measurement does). At the same time,
sensitivity of the stroboscopic variation measurement does not
depend crucially on the duration of the pumping pulses (as in the
case of the usual stroboscopic measurement), see
Fig.\,\ref{fig:4}.

Evident area of this procedure application is the laser gravitational-wave
antennae. However, in authors' opinion, it can hardly be used in traditional
topologies of gravitational-wave antennae due to two reasons. First, in this
procedure the optical energy stored in the interferometer must vary with
characteristic time $t \ll \Omega^{-1}$, where $\Omega$ is the signal
frequency. At the same time, relaxation time $\tau^*$ of the Fabry-Perot
cavities used in the large-scale gravitational-wave antennae is close to the
$\Omega^{-1}$ which makes the energy modulation very difficult from the
technological point of view.

Second, in traditional topologies of gravitational-wave antennae
very high values of the optical pumping power are required in
order to overcome the Standard Quantum Limit \cite{00p1BrGoKhTh}
which, in particular, leads to undesirable non-linear effects in
the large-scale Fabry-Perot cavities \cite{01a1BrStVy}.

In the articles \cite{96a2BrKh, 97a1BrGoKh, 98a1BrGoKh} a new class of
so-called ``intracavity'' readout schemes for gravitational-wave antennae
which allow in principle to increase sensitivity without increasing the
pumping power is proposed. In such a schemes it is necessary to measure a
small displacement of local test object with very high precision using, for
example, a small (table-top scale) interferometric position meter. We think
that stroboscopic variation measurement can be used most effectively in these
meters.

\section*{Acknowledgments}

This paper was supported in part by the California Institute of
Technology, US National Science Foundation, by the Russian
Foundation for Basic Research, and by the Russian Ministry of
Industry and Science. Special thanks to V.B.Braginsky and
S.P.Vyatchanin for fruitful discussions and valuable remarks.

\appendix
\section*{Appendix}

\section{Rectangular pumping pulses}\label{app:A}

\begin{figure}
\begin{center}
\psfragscanon \psfrag{k}[l][l][0.8]{$1-k$}
\psfrag{wt}[l]{$\omega\tau$}
\psfrag{tt=0}[rc][rc][0.85]{$\theta/\tau=0$}\psfrag{tt=0,2}[lc][lc][0.85]{$\theta/\tau=0.2$}
\psfrag{tt=0,35}[lc][lc][0.85]{$\theta/\tau=0.35$}\psfrag{tt=0,5}[lc][lc][0.85]{$\theta/\tau=0.5$}
\includegraphics[width=0.5\textwidth]{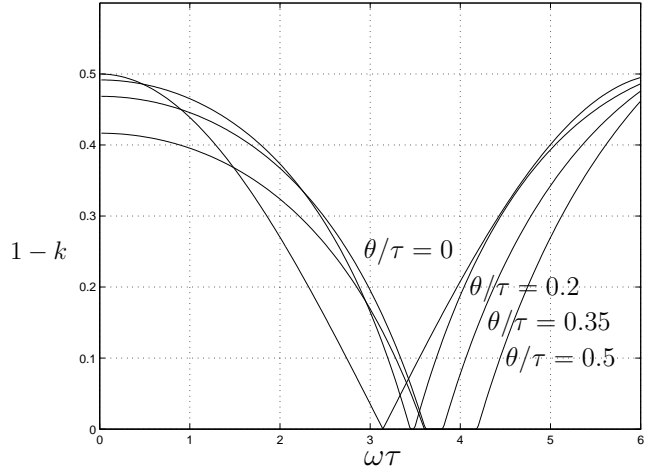}
\caption{Graphics of $1-k$ for $\theta/\tau=$0, 0.2, 0.35, and
0.5.} \label{fig:3}
\end{center}
\end{figure}

Suppose pumping pulses we have introduced above to be square
shaped. In this case function $\Delta(t)$ in (\ref{QuasiOpt1}) is equal to
\begin{equation}\label{RectDelta}
  \Delta(t)=
  \begin{cases}
    \dfrac1\theta & \text{if}\quad  |t| < \theta/2 \,, \\
    0 & \text{otherwise} \,.
  \end{cases}
\end{equation}
Taking the above assumptions and formula (\ref{DiscreteSpDensity})
into account one can obtain the following expression for the
measurement sensitivity:

\begin{equation}\label{Errk}
  \frac{S_{\rm meter}}{S_\mathrm{EQL}}
  = \frac{(\omega_m\tau/\pi)^4}{
      1 - \dfrac{32}{\omega_m\theta}\,
            \dfrac{\sin^2\frac{\omega_m\theta}{4}\,\left[
              k\cos\frac{\omega_mT}{2} + (1-k)\cos\frac{\omega_m\theta}{4}
            \right]^2}{
              \omega_m\theta
              + 2k\cos\omega_mT\sin\frac{\omega_m\theta}{2}
              + (1-k)\sin\omega_m\theta
            }
    }\,.
\end{equation}
The following values of $k$ minimize (\ref{Errk}):

\begin{equation}\label{kopt}
  k = \begin{cases}
    \dfrac{\omega_m\theta+\sin\omega_m\theta}{
      \sin\frac{\omega_m\theta}{2}(\cos\frac{\omega_m\theta}{2}-\cos\omega_mT)
    }
    - \dfrac{\cos\frac{\omega_m\theta}{4}}
        {\cos\frac{\omega_m\theta}{4}-\cos\frac{\omega_mT}{2}} \,, &
      \quad\mbox{if}\quad 2\omega_m\theta<\omega_m T<f(\theta)\,,\smallskip\\
    1\,,&\quad\mbox{if}\quad f(\theta)<\omega_mT<\pi\,, \smallskip \\
    \dfrac{\cos\frac{\omega_m\theta}{4}}
      {\cos\frac{\omega_m\theta}{4}-\cos\frac{\omega_mT}{2}} \,, &
      \quad\mbox{if}\quad \omega_mT>\pi\,.
  \end{cases}
\end{equation}
where
\begin{equation}
  f(\theta) = 2\arccos\left\{\frac14\left(
    2\cos\frac{\omega_m\theta}{4}
    - \sqrt{\frac{
        \sin\omega_m\theta-8\omega_m\theta
        + 18\sin\frac{\omega_m\theta}{2}
      }{\sin\frac{\omega_m\theta}{2}}}
  \right)\right\}
\end{equation}
Values $1-k$ for different $\theta/\tau$ are plotted in
Fig.\,\ref{fig:3}.

Corresponding values of the measurement sensitivities are defined
by the following formulae:
\begin{equation}\label{FMQuasiOptErr}
\frac{S_{\rm meter}}{S_\mathrm{EQL}}
  = \left(\frac{4}{\pi}\right)^4\,\frac{(\tau/T)^4}{
      1 + \frac{1}{6}\,(\theta/T)^2
        + \frac{29}{80}\,(\theta/T)^4
    }\,,
\end{equation}
for the free mass, and
\begin{equation}\label{OscErrk}
  \frac{S_{\rm meter}}{S_\mathrm{EQL}} =
  \begin{cases}
    \dfrac{(\omega_m\tau/\pi)^4}
      {1 - \dfrac{2}{\omega\theta}\,
              \dfrac{
                2\sin\frac{\omega_m\theta}{2}(
                  1 + 2\cos\frac{\omega_m\theta}{4}\cos\frac{\omega_mT}{2}
                )
                - \omega_m\theta
              }{
                \cos^2\frac{\omega_m\theta}{4}\,(
                  \cos\frac{\omega_m\vartheta}{4}
                  + \cos\frac{\omega_m\theta}{2}
                )^2
              }
      } \,, &\quad\mbox{if}\quad \omega_m T<f(\theta)\,, \medskip \\
    \dfrac{(\omega_m\tau/\pi)^4}
      {1 - \dfrac{32}{\omega_m\theta}\,
             \dfrac{\sin^2\frac{\omega_m\theta}{4}\cos^2\frac{\omega_mT}{2}}
               {\omega_m\theta + 2\cos\omega_mT\sin\frac{\omega_m\theta}{2}}
      } \,, &\quad\mbox{if}\quad f(\theta)<\omega_mT<\pi\,,\medskip \\
    \left(\dfrac{\omega_m\tau}{\pi}\right)^4\,, &
      \quad\mbox{if}\quad \omega_mT>\pi \,,
  \end{cases}
\end{equation}
for the harmonic oscillator.

\begin{figure}
\begin{center}
\psfragscanon \psfrag{scale}[l]{$\dfrac{S_{\rm
meter}}{S_\mathrm{EQL}}$} \psfrag{wt}[l]{$\omega\tau\rightarrow$}
\psfrag{const}[lb][lb][0.85]{${\cal
E}=\mathrm{const}$}\psfrag{units}[ll][ll]{$\dfrac{\hbar^2m^2}{4\overline
S_F\tau^5}\uparrow$}\psfrag{A}[ct][ct][0.75]{$A$}\psfrag{B}[cb][cb][0.75]{$B$}\psfrag{C}[ct][ct][0.75]{$C$}
\psfrag{tt=0}[lt][lt][0.85]{$\theta/\tau=0$}\psfrag{tt=0,2}[lt][lt][0.85]{$\theta/\tau=0.2$}
\psfrag{tt=0,35}[lt][lt][0.85]{$\theta/\tau=0.35$}\psfrag{tt=0,5}[lt][lt][0.85]{$\theta/\tau=0.5$}
\includegraphics[width=0.5\textwidth]{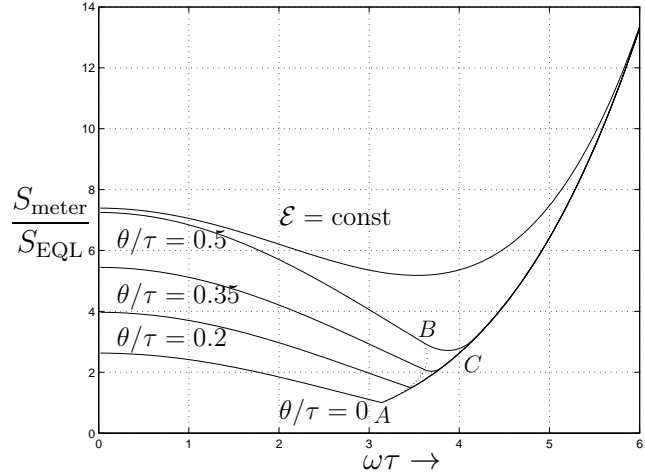}
\caption{Proposed method and DSVM procedure (${\cal
E}=\mathrm{const}$) sensitivities (noise spectral densities)
related to the EQL $\dfrac{S_{\rm meter}}{S_\mathrm{EQL}}$ when
$\theta/\tau=$0, 0.2, 0.35, and 0.5.} \label{fig:4}
\end{center}
\end{figure}

These values are plotted in Figure \ref{fig:4}. Different curves
correspond to several values of the pumping pulse duration
$\theta$. The topmost curve represents sensitivity of the DSVM
procedure with constant pumping power \cite{00a1DaKhVy}.

Here triangle \textsf{ABC} represents the ``pure stroboscopic''
area where $k=1$. It should be noted that in the area right to
this triangle the value of $S_{\rm meter}$ does not depend on
pulse duration and remains the same as in the case of
$\delta$-function like pumping (\ref{measerr2}).

\section{Energetic Quantum Limit}\label{App:B}

This appendix is devoted to the formulae (\ref{Delta_F_EQL})
derivation and also the expression for the energetic quantum limit when the
harmonic oscillator is used as a probe body.


In order to detect the external force acting on some test object
the following inequality should be fulfilled (see \cite{98a1BrGoKh}):
\begin{equation}\label{EQL}
  \int_{-\infty}^\infty |F(\omega)|^2 S_\mathrm{pos}(\omega)\,
    \frac{d\omega}{2\pi} \ge \frac{\hbar^2}{4} \,,
\end{equation}
where $F(\omega)$ is the signal force spectrum, and
$S_\mathrm{pos}(\omega)$ is the test object position fluctuations
spectral density.

Suppose position meter with back-action force spectral density
$S_F(\omega)$ is attached to the test object. Let also this object
dynamics to be described by function $D(\omega)$, that is the
operator ${\bf D}$ Fourier transform. Then spectral density
$S_\mathrm{pos}(\omega)$ can be expressed in terms of back-action
force spectral density $S_F(\omega)$ as
\begin{equation}
  S_\mathrm{pos} = \frac{S_F(\omega)}{|D(\omega)|^2} \,.
\end{equation}
Therefore, condition (\ref{EQL}) can be rewritten as:
\begin{equation}
  \int_{-\infty}^\infty \frac{|F(\omega)|^2}{S_\mathrm{EQL}(\omega)}\,
    \frac{d\omega}{2\pi} \ge 1 \,,
\end{equation}
where
\begin{equation}\label{S_EQL}
  S_\mathrm{EQL}(\omega)
  = \frac{\hbar^2}4\,\frac{|D(\omega)|^2}{S_F(\omega)} \,.
\end{equation}

Exactly the same condition can be obtained, if the uncertainty
relation for linear position meter noises is used. Really, the
total net noise of such a meter can be described by its spectral
density
\begin{equation}
  S_\mathrm{total}(\omega) = S_F(\omega) + 2D(\omega)S_{xF}(\omega) +
    |D(\omega)|^2 S_x(\omega) \,,
\end{equation}
and spectral densities of the meter noises must satisfy the following
condition:
\begin{equation}
  S_x(\omega)S_F(\omega) - S_{xF}^2(\omega) \ge \frac{\hbar^2}{4} \,.
\end{equation}
Minimizing value $S_\mathrm{total}(\omega)$ taking this additional
condition into account, one can easily show that
$S_\mathrm{total}(\omega)\ge S_\mathrm{EQL}(\omega)$.


Suppose that (i) $S_F$ does not depend on frequency\footnote{back-action
noise is supposed to be "white"} and (ii) we want to have
$S_\mathrm{EQL}(\omega)$ below some given value $S_\mathrm{EQL}$ in some
given spectral range $0 \le \omega \le \omega_\mathrm{max}$.


Let our test object be a free mass. In this case
\begin{equation}
  D(\omega) = -m\omega^2 \,,
\end{equation}
and
\begin{equation}
  S_\mathrm{EQL}(\omega) = \frac{\hbar^2m^2\omega^4}{4S_F} \,.
\end{equation}
It is evident that this spectral density is maximal if
$\omega=\omega_\mathrm{max}$, and the maximum is equal to
\begin{equation}\label{FMEQL}
  S_\mathrm{EQL} = S_\mathrm{EQL}(\omega_\mathrm{max})
  = \frac{\hbar^2m^2\omega_\mathrm{max}^4}{4S_F} \,.
\end{equation}
This is the EQL for a free mass.


Let the test object be a harmonic oscillator now. In this case we
will obtain that
\begin{equation}
  D(\omega) = m(\omega_m^2-\omega^2) \,,
\end{equation}
and
\begin{equation}\label{EQL_oscill}
  S_\mathrm{EQL}(\omega) = \frac{\hbar^2m^2(\omega_m^2-\omega^2)^2}{4S_F} \,.
\end{equation}
Suppose that we can freely tune $\omega_m$ in order to minimize
the maximum value(s) of the $S_\mathrm{EQL}(\omega)$ in the
spectral range $0 \le \omega \le \omega_\mathrm{max}$. It is
evident that in optimal case
\begin{equation}   \omega_m^2 = \frac{\omega_\mathrm{max}^2}2 \,,
\end{equation}
and maximum values of the $S_\mathrm{EQL}(\omega)$ are equal to
\begin{equation}\label{OscEQL}
  S_\mathrm{EQL} = S_\mathrm{EQL}(0) = S_\mathrm{EQL}(\omega_\mathrm{max})
  = \frac{\hbar^2m^2\omega_\mathrm{max}^4}{16S_F} \,.
\end{equation}
This is the EQL for a harmonic oscillator.

\section{The explicit optimization method}\label{App:C}
In this appendix we will pay some attention to the explicit
derivation of formulae presented above.

Here we concentrate on the expression (\ref{measerr2}) method of
optimization when conditions (\ref{bound_norm}) and (\ref{normS})
are applied. The harmonic oscillator case is considered, as the
free mass case can be obtained simply by assuming
$\omega_m\rightarrow0$.

In order to reduce the two variables problem (\ref{measerr2}) into
the single variable $v(t)$ variational problem it is necessary to
express function $S_F(t)$ in terms of $v(t)$ using the Lagrange
equation as follows:
\begin{equation}\label{LGEqn1}
 \frac{\hbar^2}{4}\frac{({\bf
  D}v(t))^2}{S_F^2}-\lambda=0\,,\,\Rightarrow\,S_F(t)=\frac{\hbar}{2\sqrt{\lambda}}|{\bf D}v(t)|\,,
\end{equation}
where $\lambda$ is the Lagrange multiplier. There is no difficulty
now to transform the initial optimization problem into the
Liapunov's one (See \cite{Galeev,Tikhomirov} for details) by
changing variable $v(t)$ in accordance with the formula ${\bf
D}v(t)=u(t)$:
\begin{equation}\label{LiapPr}
\begin{array}{cc}
 \int_{-\tau/2}^{\tau/2}|u(t)|\,dt\rightarrow\min,\quad
  \int_{-\tau/2}^{\tau/2}\,u(t)\,dt=\omega_m^2,\\
\int_{-\tau/2}^{\tau/2}\sin(\omega_m(\tau/2-t))\,u(t)\,dt=0,\quad
\int_{-\tau/2}^{\tau/2}\cos(\omega_m(\tau/2-t))\,u(t)\,dt=0.
\end{array}
\end{equation}
Using the ideas represented in \cite{Galeev,Tikhomirov} the
following minimum value of (\ref{LiapPr}) can be obtained 

\begin{equation}\label{UndefSol}
u(t)=
 \begin{cases}
\mu\delta(t-\frac{\tau}{2})+\nu\delta(t)+\xi\delta(t+\frac{\tau}{2}),\quad\mbox{при}\quad\omega_m\tau\in[0,\,\pi]\\
\beta\delta(t-\vartheta)+\gamma\delta(t+\vartheta)
\quad\mbox{при}\quad\omega_m\tau\in(\pi,\,\infty),
\end{cases}
\end{equation}
where $\{\mu,\,\nu,\,\xi\}$ and $\{\beta,\,\gamma,\,\vartheta\}$
are indefinite coefficients that can be defined through
substitution of function $u(t)$ into the supplementary condition
equations in (\ref{LiapPr}). The final expression for $u(t)$ may
be represented as
\begin{equation}\label{OscSolution}
 u(t)=
 \begin{cases}
\dfrac{\omega_m^2}{1-\cos\frac{\omega_m\tau}{2}}\left[\delta(t-\frac{\tau}{2})-2\cos\frac{\omega_m\tau}{2}\delta(t)+\delta(t+\frac{\tau}{2})\right],\quad\mbox{при}\quad\omega_m\tau\in[0,\,\pi]\\
\dfrac{\omega_m^2}{2}\left[\delta(t-\frac{\pi}{2\omega_m})+\delta(t+\frac{\pi}{2\omega_m})\right],\quad\mbox{при}\quad\omega_m\tau\in(\pi,\,\infty).
 \end{cases}
\end{equation}
Filtering function $v(t)$ is expressed in terms of $u(t)$ as
\begin{equation*}
  v(t)=\frac{1}{\omega_m}\int_{-\frac{\tau}{2}}^t\sin(\omega_m(t-t'))u(t')\,dt'\,.
\end{equation*}


\begin{thebibliography}{10}

\bibitem{67a1eBr}
{V.B.Braginsky},
\newblock Sov.\,Phys.\,JETP {\bf 26}, 831 (1968).

\bibitem{00a1BrGoKhTh}
{V.B.Braginsky, M.L.Gorodetsky F.Ya.Khalili and K.S.Thorne},
\newblock Physical Review D {\bf 61}, 4002 (2000).

\bibitem{300years}
{K.S.Thorne},
\newblock {\em 300 Years of Gravitation},
\newblock Cambridge University Press, 1987.

\bibitem{78a1eBrKhVo}
{V.B.Braginsky, Yu.I.Vorontsov, F.Ya.Khalili},
\newblock Sov.\,Phys.\,JETP-Lett. {\bf 33}, 405 (1978).

\bibitem{95a1VyZu}
{S. P. Vyatchanin ш E. A. Zubova},
\newblock Physics Letters A {\bf 201}, 269 (1995).

\bibitem{98a1Vy}
{S. P. Vyatchanin},
\newblock Physics Letters A {\bf 239}, 201 (1998).

\bibitem{02a1KiLeMaThVy}
{H.J.Kimble, Yu.Levin, A.B.Matsko, K.S.Thorne and S.P.Vyatchanin},
\newblock Physical Review D {\bf 65}, 022002 (2002).

\bibitem{00a1DaKhVy}
{S. L. Danilishin, F. Ya. Khalili and S. P. Vyatchanin},
\newblock Physics Letters A {\bf 278}, 123 (2000).

\bibitem{92BookBrKh}
{V.B.Braginsky, F.Ya.Khalili},
\newblock {\em Quantum Measurement},
\newblock Cambridge University Press, 1992.

\bibitem{00p1BrGoKhTh}
{V.B.Braginsky, M.L.Gorodetsky, F.Ya.Khalili and K.S.Thorne},
\newblock Energetic quantum limit in large-scale interferometers,
\newblock in {\em Gravitational waves. Third Edoardo Amaldi Conference,
  Pasadena, California 12-16 July, ed. S.Meshkov, Melville NY:AIP Conf. Proc.
  523}, pages 180--189, 2000.

\bibitem{Likbez2001}
{V.B.Braginsky, M.L.Gorodetsky, F.Ya.Khalili, A.B. Matsko, K.S.Thorne,
  S.P.Vyatchanin},
\newblock LANL preprint {\bf gr-qc/0109003}, submitted to Phys. Rev. D (2001).

\bibitem{Galeev}
{E. M. Galeev, V. M. Tikhomirov},
\newblock {\em The brief course of optimization problems},
\newblock The MSU Publishing House, 1989.

\bibitem{01a1BrStVy}
{V.B.Braginsky S.E.Strigin and S.P.Vyatchanin},
\newblock Physics Letters A {\bf 287}, 331 (2001).

\bibitem{96a2BrKh}
{V. B. Braginsky, F. Ya. Khalili},
\newblock Physics Letters A {\bf 218}, 167 (1996).

\bibitem{97a1BrGoKh}
{V.B.Braginsky, M.L.Gorodetsky, F.Ya.Khalili},
\newblock Physics Letters A {\bf 232}, 340 (1997).

\bibitem{98a1BrGoKh}
{V.B.Braginsky, M.L.Gorodetsky, F.Ya.Khalili},
\newblock Physics Letters A {\bf 246}, 485 (1998).

\bibitem{Tikhomirov}
{V. M. Alekseev, V. M. Tikhomirov, S. V. Fomin},
\newblock {\em The optimal control},
\newblock Science, 1979.

\end{thebibliography}

\end{document}